# $CO_2$ HYDRATE DISSOCIATION AT LOW TEMPERATURES – FORMATION AND ANNEALING OF ICE Ic

**Andrzej Falenty, Werner F. Kuhs**[*]
**Geowissenschaftliches Zentrum der Universität Göttingen**
**Abteilung Kristallographie,**
**Goldschmidtstraße 1, 37077 Göttingen**
**GERMANY**

**Thomas C. Hansen**
**Institut Max von Laue-Paul Langevin**
**6 rue Jules Horowitz, BP 156, 38042 Grenoble Cedex 9**
**FRANCE**

**ABSTRACT**
Dissociation of gas hydrates below 240 K leads to the formation of a metastable form of water ice, so called cubic ice (Ic). Through its defective nature and small particle size the surface film composed of such material is incapable of creating any significant diffusion barrier. Above 160 K, cubic ice gradually transforms to the stable hexagonal (Ih) form on laboratory time scales. The annealing, coupled with a parallel decomposition of gas hydrates, accelerates as temperature rises but already above 190 K the first process prevails, transforming cubic stacking sequences into ordinary Ih ice within a few minutes. Remaining stacking faults are removed through very slow isothermal annealing or after heating up above 240 K. The role of the proportion of cubic stacking on the decomposition rate is discussed.

A better understanding of the dissociation kinetics at low temperatures is particularly important for the critical evaluation of existing hypotheses that consider clathrates as a potential medium that actively participate in geological processes or is able to store gases (e.g. $CH_4$, $CO_2$ or Xe) in environments like comets, icy moons (i. e. Titan, Europa, Enceladus) or on Mars. Here, we present kinetics studies on the dissociation of $CO_2$ clathrates at isothermal and isobaric conditions between 170 and 190K and mean Martian surface pressure. We place special attention to the formed ice and demonstrate its influence on the dissociation rates with a combination of neutron diffraction studies (performed on D20 at ILL/Grenoble) and cryo-SEM. More detailed crystallographic information has been acquired via a flexible stacking-fault model capable of revealing the time evolution of the defect structure of ice Ic in terms of stacking probabilities and crystal size.

**INTRODUCTION**
Although it is well known that gas hydrates below the freezing point of water transform to gas and water ice [1], its influence on the dissociation kinetics causes still confusion in spite of a few decades of research.

A great effort to explore this puzzle has been dedicated to a temperature region between the melting point of ice and 240 K, where anomalously slow decomposition rates have been observed [2-6]. The kinetic anomaly extending across a broad temperature and pressure field, the so-called

---
[*] Corresponding author: Phone: +49 551 39 3891 Fax: +49 551 39 9521 E-mail: wkuhs1@gwdg.de

"self-preservation" phenomenon, was closely linked to a growth of hexagonal ice (ice Ih) on the surface of decaying gas hydrates and their subsequent fairly rapid coarsening [7]. With decreasing temperatures the kinetic anomaly weakens due to the slower annealing rates and accumulation of defects by ice crystals [6, 7].

Recent diffraction studies prove that water molecules released upon the dissociation of clathrates at even lower temperatures, below 240 K in fact do not crystallize in a hexagonal lattice. Instead, so-called "cubic ice" (ice Ic) is formed [7]. Although this polymorph of water ice is a fairly common product of a number of low temperature processes [8, 9 and references within], its influence on the dissociation kinetics of clathrate hydrates phase has been not closely explored. A better understanding of the role of ice Ic on the decomposition rates may shed new light on the ability of clathrates to persist in exotic, extraterrestrial environments at temperatures well below the melting point of ice. Potential locations where such processes might take place could be comets [10], icy moons [11-13] or planets like Mars [14 and references within, 15-17] where clathrate hydrates were postulated to exist in substantial amounts. In this context, it is particularly interesting whether an active participation in various geo(morpho)-logical processes or any passive ability to store gases (e.g. $CH_4$, $CO_2$ or Xe) is conceivable. Without the knowledge of the low temperature decomposition kinetics of gas hydrates many hypotheses based on their assumed occurrence will remain pure speculation.

Diffraction studies on "cubic ice" of various origins [8, 9] show this phase to consist of small particles build from a mixture of cubic and hexagonal stacking sequences along the *c* axis. This can be readily recognized by the considerable broadening of reflections belonging to ideal cubic ice, and in particular in a well pronounced skewness of the 111 and 222 reflections [8, 9, 18, 19]. The typical particle size calculated from the Scherrer formula is a few hundred Å with a maximum radius not exceeding 300 Å [18, 20, 21]. Additional supportive arguments for these observations were found by cryo-SEM imagery where well recognizable "kinks" on prismatic planes (oriented parallel to the c-axis) were commonly observed on ice particles of a few μm in diameter [7].

An ice film composed of these particles, formed on the surface of decaying clathrate hydrates, is unlikely to form any appreciable diffusion barrier. Gas molecules escaping from collapsing cages can use numerous grain boundaries and defective zones as fairly rapid transportation pathways. Yet, this picture is disturbed by the fact that cubic ice requires low temperatures (below ~ 150 - 160 K) to be stable on a laboratory timescale [7, 22]. Above this temperature, ice Ic is metastable and gradually transforms to the thermodynamically favoured hexagonal structure, thus limiting the number of possible diffusion pathways. The rate of change quickly increases towards higher temperatures and already from about 190 - 200 K, the annealing transforms the majority of cubic sequences on a time scale of a few minutes. The remaining content of cubic sequences, however, persists for prolonged time or until heating above 230 – 240 K where the transformation to an ideal hexagonal structure becomes rapidly complete [7].

Here, we present progress in the understanding of the influence of Ic on the dissociation rates of $CO_2$ clathrates at conditions relevant to Martian polar and sub-polar surface conditions; here we present work done in temperature range from 167.7 to185 K in a series of isobaric and isothermal *in-situ* neutron diffraction experiments complemented by *ex-situ* cryo-SEM imaging. A detailed ice Ic analysis in terms of the evolution of particle size and stacking fault as a function of time has been performed with a flexible stacking-fault model developed on and applied to cubic ice obtained from high-pressure polymorphs of ice [8, 9].

## SAMPLE PREPARATION AND EXPERIMENTAL METHODS

### Starting material

Deuterated $CO_2$ clathrate used in this experiment was prepared from ice spheres formed by spraying $D_2O$ water into liquid $N_2$ [6]. Recovered $CO_2$ clathrates were later crushed under liquid $N_2$ and sieved through a set of 200 and 300 μm meshes. In order to prevent a possible contamination with $H_2O$ frost the procedure has been performed in a sealed glove box under an inert $N_2$ atmosphere. SEM images of the clathrate hydrate powders show irregular particles characteristic for crushed material (Figure 1).

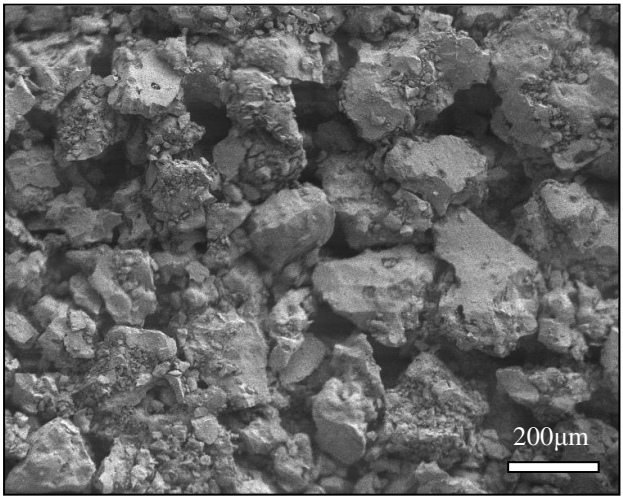

Figure 1. 200 - 300 µm size $CO_2$ clathrate hydrate powder. On some larger particles one can notice fine detritus electrostatically attached during the crushing.

**Analytical methods**
In-situ neutron diffraction experiments were performed on the high-flux 2-axis neutron diffractometer D20 at the Institut Laue-Langevin (ILL), Grenoble, France [23]. The diffractometer for these experiments was configured for a fast data acquisition on the expense of some resolution. Data was gathered with a time step of 60 or 300 s depending on the decomposition rate. A PID loop temperature controller attached to an "orange" He-flow cryostat operating between 1.7 and 300 K has provided temperature control. A detailed sample insertion procedure can be found elsewhere [6]. Quantitative information on the $CO_2$ clathrate content has been obtained with a full pattern Rietveld refinement package – GSAS [24] – for every time frame. Due to the disordered nature of cubic ice, the peak shapes of its diffraction pattern cannot be handled properly by a standard Rietveld refinement; ice reflection positions were therefore excluded from the pattern for an initial treatment. Likewise, aluminium reflections and other features (precipitates) from the pressure cell were excluded. The atomic positions and displacement parameters for $CO_2$-hydrate were kept fixed. Refined parameters were limited to the zero-shift, overall scale factor and three to four profile and six to eight Chebychev type background parameters. The weight fraction of the clathrate phase, $\alpha$, is then obtained as a function of time from the scaling factor. An approximate degree of perfection of ice formed has been retrieved for every time step by determining the intensity ratio of the 002 and 100 ice Ih Bragg peaks ($I_{002}/I_{100}$) using the LAMP software [25].

The starting material and the samples after the neutron diffraction runs were investigated with an *ex-situ* cryo-SEM using a FEI Quanta 200F and a LEO 1530 Gemini instrument equipped with Polaron and Oxford CTH1500HF cryo-stages, respectively. Uncoated, partially decomposed clathrate powders were studied at about 90 K (with liquid $N_2$ as coolant) and a pressure of about 0.1 Pa. In order to minimize sample deterioration in the electron beam, a fairly low acceleration voltage of 2.5 keV was used.

**Ic modelling**
The recently developed flexible stacking-fault model [8, 9] is capable of revealing the time evolution of the defect structure of ice Ic in terms of stacking probabilities. Here, we processed datasets from four isothermal runs at 167.7, 170, 175, 180 and 185 K performed at the constant pressure of 0.6 kPa. The experimental runs cover the temperature range where the highest accumulation of cubic stacking sequences is deduced from ($I_{002}/I_{100}$) Ih Bragg peaks ratio. Guided by the earlier work [9] the following parameters have been refined: scaling factors of ice Ic and $CO_2$ clathrate, and lattice parameters $a$ and $c$ of a trigonal unit cell in which ice Ic is modelled. The anisotropic size broadening parameter $a_{00}$ for cubic ice and the lattice parameter $a$ of the primitive cubic unit cell of the clathrate were refined first for every temperature and kept fixed in the remaining isothermal series of data acquisitions. In the case of the 175 and 185 K data it was necessary also to refine a scaling factor for ice Ih due to a presence of a small amount of unreacted cores of ice spheres used as a starting material. $a$ and $c$ lattice parameter for Ih ice for respective temperatures were taken from the literature [26]. The number of stacked layers $N_c$ was freely refined, but below $N_c \approx 150$, an increased scattering of this parameter has been observed mostly due to limited resolution of the D20 diffractometer in the chosen configuration. Finally, we refined the stacking probabilities $\alpha$, $\beta$, $\gamma$ and $\delta$ [8]. The latter set of four parameters describe the probability for two layers obeying the stacking rules HH, HK, KH and KK[1], respectively, to be followed by a layer

---
[1] A stacking obeying a hexagonal (H) stacking rule (a new layer $n+1$ is identical to the layer $n-1$ be-

following the stacking rule K. All four stacking probabilities are constrained between 0 and 1. In ideal cubic ice the fourth parameter ($\delta$) is equal to 1 (and the others are greater than zero). The ideal hexagonal lattice forbids any occurrence of K layers, thus, the first probability ($\alpha$) is zero (and the others smaller than one). Zero means that after two stacking rules (HH in the case of alpha) the probability of a cubic one following is zero, thus, the probability of hexagonal stacking is one. For a perfect hexagonal stacking, one does not need all four probabilities to be zero, only alpha: provided the others are different from one, the probability of HH appearing is not zero, and once that sequence appeared, all further ones are necessarily hexagonal. From $\alpha$, $\beta$, $\gamma$ and $\delta$ determined for every single diffraction pattern the pair-occurrence probabilities $w_{HH}$, $w_{KK}$, $w_{HK}$, and $w_{KH}$ were calculated. With $w_{HK} = w_{KH}$, an evolution of the proportion of cubic sequences over time can be presented as "cubicity" ($w_{KK} + w_{HK}$), which gives the probability that a layer sequence is obeying cubic stacking rules. Due to convergence problems at very low Ic weight fractions, it was necessary to perform sequential refinements starting from the last pattern of a series of data sets. A broad background feature coming from the pressure cell had a destabilizing effect on the refinement, too. The latter problem could be largely reduced by a background subtraction. For the runs at 170 and 180 K the background description was achieved with a fitted polynomial curve. In case of the experiments performed at 175 and 185 K, measured during another experimental campaign, we used a pattern of an empty cell instead. Due to some texture in the aluminium alloy of the cell, in both cases, parts of the pattern containing Al-peaks still had to be excluded.

**RESULTS AND DISCUSION**

The results presented here were obtained during two experimental campaigns designed to probe dissociation rates of $CO_2$ clathrates at $p$-$T$ conditions relevant to the Martian polar surface and close subsurface (Figure 2). All runs have been performed at a constant temperature and a constant

---

neath the layer $n$) followed by another one is called HH; one followed by a cubic (K) stacking rule (a new layer $n+1$ is not identical to the two previous layers $n$ and $n$-1) is called HK, and so forth; $n$ is here an arbitrary index of a particular layer, stacked in the direction of the $c$ axis.

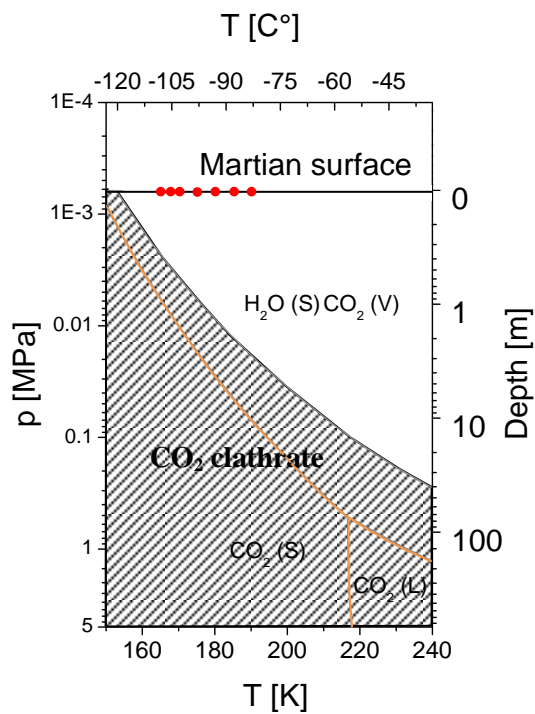

Figure 2. Low temperature fragment of the $CO_2$ clathrate stability field (Shaded area) relevant to the Martian $p$-$T$ conditions. Red points on the solid black line mark dissociation conditions for the experiments shown here. Orange line marks stability boundaries for $CO_2$ phases.

pressure of 0.6 kPa (later on, only temperatures are used to distinguish the kinetic curves).

*In-situ* **neutron diffraction experiments**

At the lowest temperature (165 K) outside the thermodynamic stability field of $CO_2$-hydrate (boundary at about 2.3 kPa), dissociation was not observed within nearly 4 h (Figure 3), although D20 is capable to resolve phases with as little as 0.1 weight %. The reason for this behaviour lays most likely in a very limited mobility of water molecules at such a low temperature. This efficiently hampers the transformation of unstable $CO_2$ clathrates to the thermodynamically favoured ice and $CO_2$ gas. Additionally, since this observation is based on a general property of water ice, we speculate that a similar behaviour might be expected also for clathrate hydrates of the same structure hosting other gas species. Moreover, we think that excursions of several hours from the stability field up to this temperature are likely to leave a potential clathrate unaffected. It remains an open question whether a very sluggish transformation well exceeding the laboratory timescales takes place. The sample destabilized at 165 K was brought back into the stability field and after about 1 h was destabilized again at 167.7 K where a sluggish dissociation could be observed. The following experiments performed at successively higher temperatures with fresh samples show a general increase of the dissociation rate but already at 175 K, and also at higher temperatures, an anomalous behaviour has been observed: after an initial fairly rapid dissociation, the reaction visibly slows down. This unusual behaviour can be linked to a rapid transformation of ice Ic into Ih that is known to exist in this temperature range [7]. A confirmation of this assumption has been found in the intensity ratio $I_{002}/I_{100}$ between the 002 and 100 Bragg peaks (referring to the hexagonal indexing of ice Ih), which are particularly sensitive to this transition [7]. Yet, bearing in mind that a peak shape profile can be also affected by other factors (e.g., anisotropic particle size and correlations in the disordered stacking), this method should be treated only as a rough approximation. The initial maximum of the $I_{002}/I_{100}$ ratio over time successively diminishes with rising temperature but even at the highest temperature studied (190 K) it does not fall below 1.5 (in perfect ice Ih, the ratio is about 0.55). The first two experiments (167.7 and 170 K) show a slowly rising trend until a steady state between the accumulation of fresh Ic and the annealing to Ih is achieved. At 175 K, a similar behaviour is observed but the slower decomposition works in favour of the transformation to Ih.

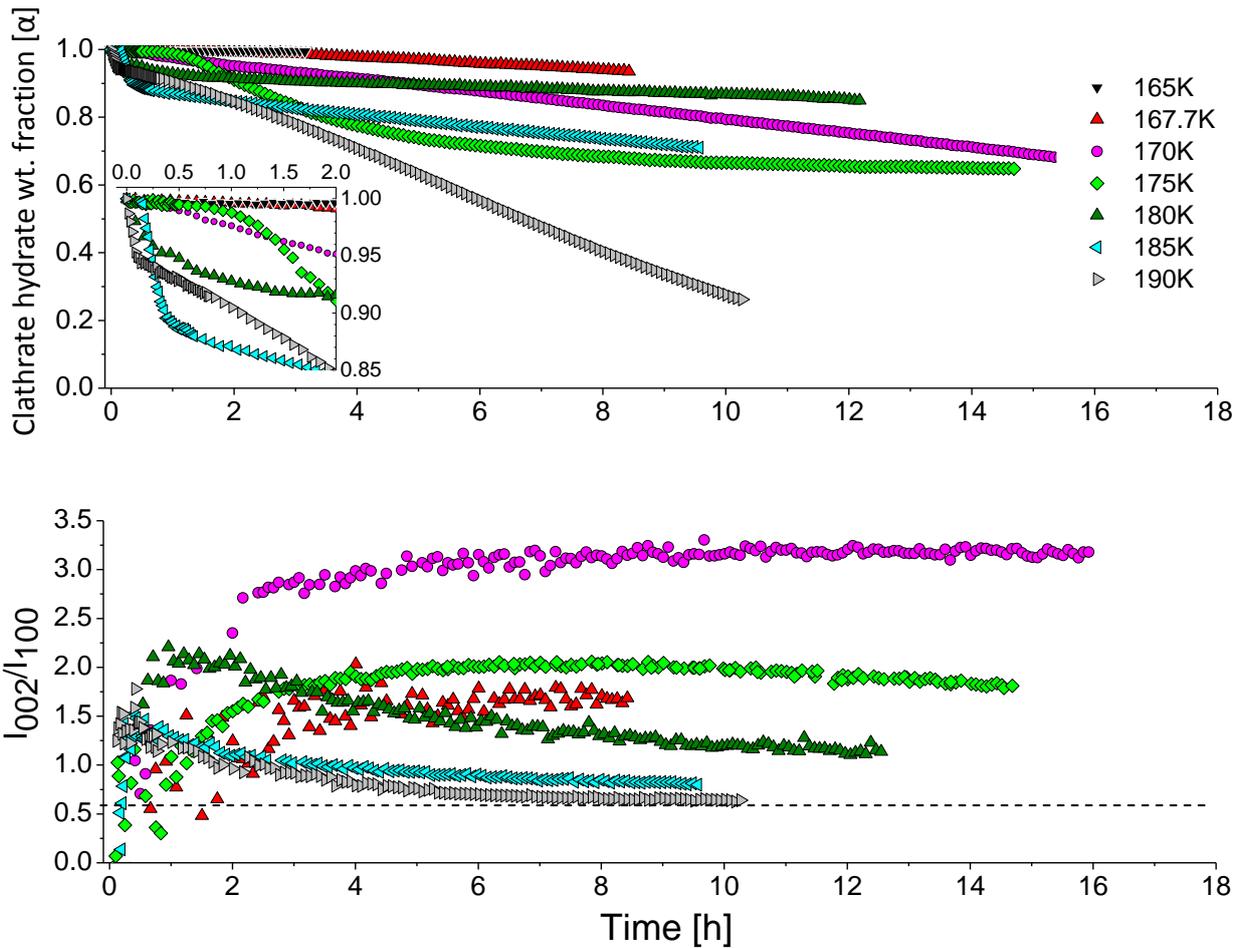

Figure 3. Dissociation curves (above) performed on ~ 250 μm $CO_2$ hydrate powders at the isobaric (0.6 kPa) conditions for selected temperatures. On the blow up one can see a well recognizable rapid change in the dissociation rate during the first few hours of the reaction. Below, the evolution of the $I_{002}/I_{100}$ ratio of Ih Bragg peaks. A dashed line marks a ratio for the ideal hexagonal ice. Color, as well as symbol code for respective experiments is valid for both plots.

From 180 K to 190 K, a sharp the rising $I_{002}/I_{100}$ ratio is followed by more and more rapid decrease towards the ideal Ih value of 0.55 as the annealing accelerates with temperature. Even so, a minor fraction of cubic stacking is still observable with the $I_{002}/I_{100}$ ratio slightly above the ideal value for ice Ih at the terminal point for each experiment (after several hours).

**Cryo-SEM observations**

Even if neutron diffraction is capable of delivering a wealth of bulk data, information on the local scale is beyond its reach. Cryo-SEM can fill this gap offering a valuable view on the microstructure of the ice film developing on the surface of a decaying clathrate.

At the lowest temperature run (167.7 K), which was halted after 8 h, the coverage is incomplete and the clathrate surface is still well visible (Figure 4A). Fairly well developed ice crystals with well recognizable prismatic and basal planes cover clathrate particles. Their average diameter is 4–6 μm. Such an appearance is in accordance with a low driving force that leads mainly to a 3D growth of ice. We have also found multiple bright and dark striation patterns ("kinks") perpendicular to the c-axis, which are characteristic for stacking faults. This is in accordance with the $I_{002}/I_{100}$ ratio presented above. At 175 K, the ice film is composed of two populations of ice crystals: Larger ones of 6 to 10 μm in diameter and smaller ones of 2 to 3 μm, which fill the free space between them (Figure 4B). Also, here, one can see numerous kinks on prismatic planes that are characteristic for stacking faults. After nearly 15 h of reaction, one can also start to notice signs of a slowly progressing annealing, which leads to a rounding of sharp edges and a smoothening of the ice film. Along with this process the ice barrier becomes more and more efficient in slowing down further decomposition. This process is even better visible on clathrate particles, decomposed at 180 K for 12 h, where rounded ice crystals of about 3 μm in diameter completely cover the surface (Figure 4C). Additionally, here, the dissociation rate considerably drops down due to the increasing diffusion limitation caused by ice. Although the $I_{002}/I_{100}$ ratio at the end of the 180 K run suggests a still considerable defectiveness of the accumulated ice, we were unable to localize any well recognizable striation patterns or kinks.

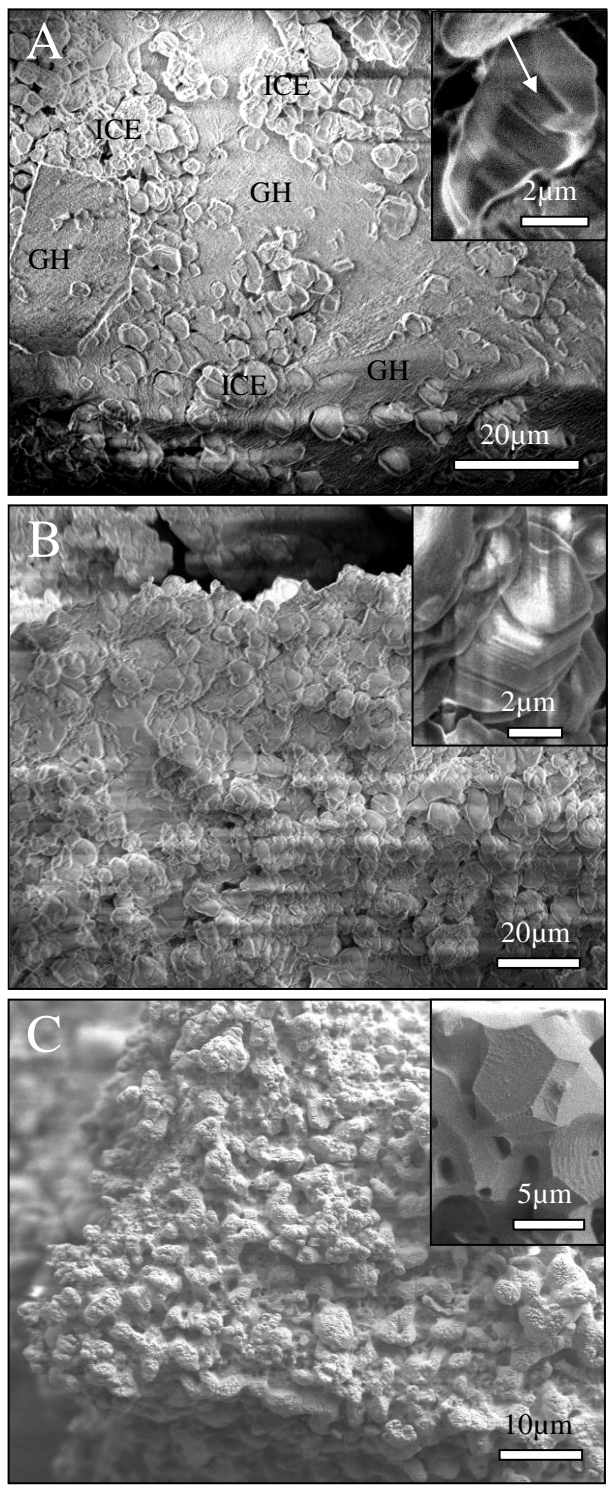

Figure 4. Cryo-SEM images of ice films observed on $CO_2$ clathrates recovered after the neutron diffraction runs at three selected temperatures: A) After 8 h at 167.7 K, gas hydrate (GH) particles are sparsely covered with stacking faulty ice crystals (magnification in the upper right corner), B) A full, dense ice coating formed after 14.5 h at 175 K. Also here, one can find evidence for defective crystals (blow-up), C) Partially annealed ice crystals observed on the sample recovered after 12 h at 180 K.

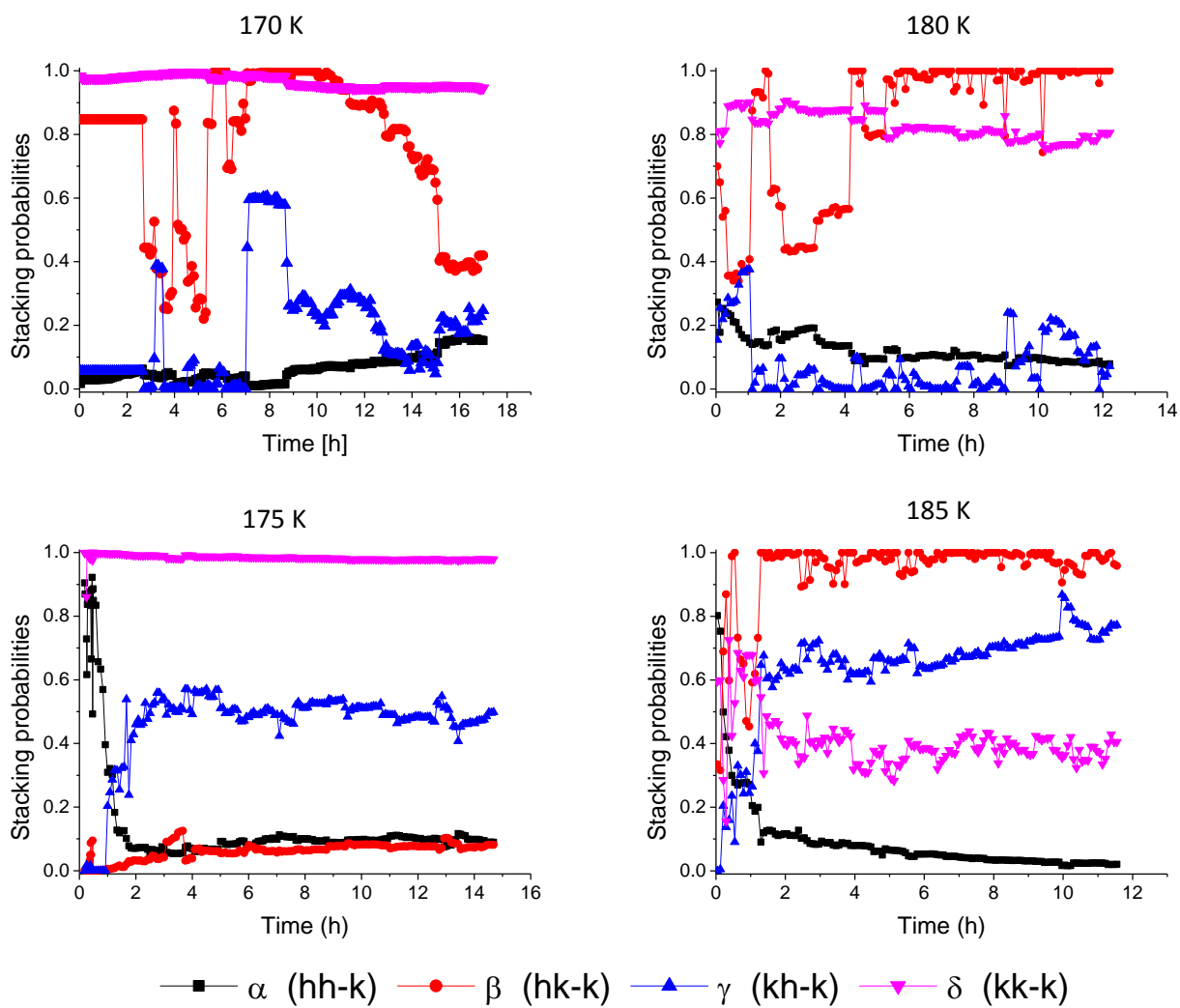

Figure 5: Evolution of stacking probabilities α, β, γ and δ for decomposition experiments at 170, 175, 180 and 185K refined with the stacking fault model.

**Stacking fault modeling – "cubicity"**
Here we report on ongoing modelling process that offer a more detailed view on the formation and annealing process of "cubic ice". Calculations were demanding in terms of computing power and prone to instabilities for low amounts of ice present, which can be well seen in the scattering of the initial data points (Figure 5). There are strong correlations between the four stacking probabilities, in particular β and γ. Even though further data treatment may be needed, we are able to make some first observations.

At the lowest processed temperature presented here (170 K) from the initial few hours of the reaction the high values for the probability δ show a strong tendency to form cubic ice sequences. This is in accordance with a rapid increase of the $I_{002}/I_{100}$ ratio (Figure 3). Later, after approximately 6 h, a progressing annealing becomes more apparent (decreasing δ). This transition is concomitant to the clear flattening the $I_{002}/I_{100}$ ratio. Interestingly, the dissociation speed does not seem to be affected by this process. The reason for this observation is not straightforward and may be found in a complex interaction between a high fraction of cubic stacking in ice, close to 65% (Figure 6) and the small crystallite sizes inferred through the number of stacked layers $N_c$ (Figure 7). In particular the importance of the later mechanism has been shown in our earlier studies [6] where specific arrangements and appearances of individual crystals can indeed dramatically influence the kinetics. At 175 K too, cubic sequences dominate in the ice crystals (high $\delta$). The maximum amount of cubic sequences calculated for this temperature reaches nearly 50% (Figure 6) with a falling trend after eight hours. $N_c$ for this run increases, at a smaller rate in comparison to the 170 K case, yet the overall size at a given time is higher as one would expect (Figure 7). Although the modelling shows a rather high defectiveness of the formed cubic ice, its presence does not explain why the clathrate dissociation slows down to such extent (Figure 3). We suspect that in this particular case the microstructure of ice (here indeed on a micron-scale) understood as the mutual three-dimensional arrangements of growing crystals may play a pivotal role. Similar to our earlier studies at higher temperatures, a tight, dense coating observed under SEM (Figure 4B) arising from the ongoing annealing may well be capable of creating an efficient barrier for gas molecules, in spite of the high content of cubic stacking sequences. The experiment at 180 K brings a significant change in the stacking probabilities. Long successions of cubic sequences, which were dominating at lower temperatures, become gradually less frequent (falling $\delta$). At this temperature the value decreases from about 50% initially to 30% (Figure 6). A similar trend can also be observed in the $I_{002}/I_{100}$ ratio. At the highest processed temperature, 185 K, fresh cubic ice is still rapidly formed (high $\beta$) but decaying $\delta$ suggests that longer sequences of cubic ice become less frequent over time. Initially, the cubicity still reaches the level of the 180 K run, but later quickly decays to ~15% (Figure 6). Higher temperature also favours $N_c$, which increases more rapidly and exceeds 300 layers already after 7 h.

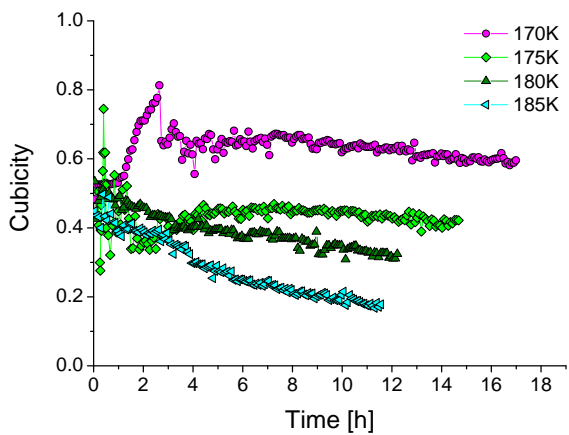
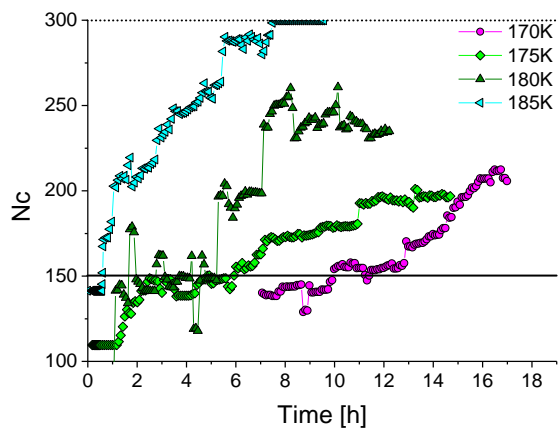

Figure 6: Time evolution of the "cubicity" of ice formed upon the decomposition of $CO_2$ hydrates for selected temperatures. Some stronger scattering at the initial part of the reactions is caused by decreased stability of the model due to a low overall fraction of cubic ice in the sample

Figure 7: Growth of ice crystal expressed through the number of stacked layers ($N_c$). The solid line corresponds to the lower resolution limit of the diffractometer; below this level the refinement is still possible but in cost of increased error. The dotted line sets the upper measurable level for the used setup.

The complex nature of the largely concomitant processes of ice Ic formation (by gas hydrate decomposition and gas release), the transformation of ice Ic into ice Ih (by local changes in their stacking sequences or ice Ih nucleation) and the coarsening of the ice crystals (driven by lowering the surface energy), all happening with different time constants, cannot be considered to be fully understood. Moreover, a full understanding on a nanoscopic or molecular scale, accessible by diffraction experiments, will be insufficient to explain the decomposition process of gas hydrates at low temperature. Referring only to nanoscopic pathways for the escape of gas molecules (provided by grain boundaries, whose density is calculated from the Scherrer estimate of particle size,, or along lattice imperfections like dislocation lines possibly linked to stacking faults) is an oversimplification. The arrangement of ice crystallites on a micron-scale (largely depending on the loca-tion of nucleation and the directional aspect of anisotropic ice crystal growth) as seen by cryo-SEM, will also influence the decomposition process by providing pathways along pores for the out-diffusing gas molecules. Only the combination of cryo-SEM and diffraction experiments can give us a much better insight into these processes; a first step in this direction was presented here. More work is undoubtedly needed for a full understanding.

**SUMMARY**

We used a combination of neutron diffraction and *ex situ* cryo-SEM observations to explore the evolution of ice formed upon decomposition of $CO_2$ clathrates at temperatures from 165 to 190 K.

At 165 K the very low mobility of water molecules efficiently slows down the dissociation and thus the transformation to cubic ice. At higher temperatures we observe an accumulation of metastable cubic ice and its concomitant annealing to the thermodynamically favoured hexagonal ice. The first process seems to be closely related to the dissociation rate, and thus the *p-T* conditions, while the second one is purely temperature driven. More detailed information on the stacking sequences, lattice parameters and particle sizes has been obtained through a refinement capable of dealing with a stacking faulty ice. We have good evidence that cubic ice does not create an efficient shielding capable of retarding the dissociation of gas hydrates but it important to point out that it is only one of the many parameters influencing the dissociation rates. Modifications of the decomposition process are expected from the arrangement of ice crystals and their appearance (3D, 2D plates, dendrites and combination of those) that can reduce or increase the permeability of the ice film. On a longer time scale, the defective nature and small particle size (i.e. large grain boundary surface) of cubic ice is likely to provide additional diffusion pathways for escaping gas molecules thus supporting the dissociation process.

In the light of our experiments we see difficulties in preserving finely dispersed gas hydrates (e.g. when grown from water frost), once outside their field of stability in the investigated temperature region. Nevertheless, larger particles might still be capable of developing an ice film thick enough to preserve some portion of existing gas hydrates for prolonged times. These findings are relevant in an often-discussed astrophysical and planetological context.

**ACKNOWLEDGMENTS**
We thank the Deutsche Forschungsgemeinschaft for financial support (Grant Ku 920/11), the Institut Laue-Langevin (ILL) in Grenoble for beam time and support and Kirsten Techmer (Göttingen) for her help in the cryo-SEM sessions.